\documentclass[conference]{IEEEtran}
\usepackage{cite}
\usepackage{amsmath,amssymb,amsfonts}
\usepackage{algorithmic}
\usepackage{graphicx}
\usepackage{textcomp}
\usepackage{subfig}
\usepackage{subcaption}
\usepackage{xcolor}
\usepackage{fancyhdr}
\def\BibTeX{{\rm B\kern-.05em{\sc i\kern-.025em b}\kern-.08em
    T\kern-.1667em\lower.7ex\hbox{E}\kern-.125emX}}
\begin{document}


\title{AUV-Assisted Underwater 6G: Environmental Modeling and Multi-Stage Optimization}

\author{
    \IEEEauthorblockN{Mustafa Yavuz Engin\IEEEauthorrefmark{1}, Mehmet Ozdem\IEEEauthorrefmark{2}, and Tuğçe BILEN\IEEEauthorrefmark{3}}
    
    \IEEEauthorblockA{\IEEEauthorrefmark{1}Department of Computer Engineering\\}
     \IEEEauthorblockA{\IEEEauthorrefmark{2}Turk Telekom, Istanbul,Turkey }
    \IEEEauthorblockA{\IEEEauthorrefmark{3}Department of Artificial Intelligence and Data Engineering\\Faculty of Computer and Informatics\\
    Istanbul Technical University, Istanbul, Turkey\\
    Email: enginm20@itu.edu.tr, mehmet.ozdem@turktelekom.com.tr, bilent@itu.edu.tr}
}

\maketitle

\begin{abstract}
6G communication plays a crucial role in enabling high-speed, low-latency data transfer for underwater operations. Underwater communications need optimization support to overcome various problems such as packet loss and latency.
This study presents a simulation model for underwater 6G networks, focusing on the optimized placement of sensors, AUVs, and hubs. The network architecture consists of fixed hub stations, mobile autonomous underwater vehicles (AUVs), and numerous sensor nodes. Environmental parameters such as temperature, salinity, and conductivity are considered in the transmission of electromagnetic signals; signal attenuation and transmission delays are calculated based on physical models. The optimization process begins with K-Means clustering, followed by sequential application of Genetic Algorithm (GA) and Particle Swarm Optimization (PSO) to refine the cluster configurations. The simulation includes key network dynamics such as multi-hop data transmission, cluster leader selection, queue management, and traffic load balancing. To compare performance, two distinct scenarios—one with cluster leaders and one without—are modeled and visualized through a PyQt5-based real-time graphical interface. The results demonstrate that 6G network architectures in underwater environments can be effectively modeled and optimized by incorporating environmental conditions.
\end{abstract}


\thispagestyle{fancy}

\pagestyle{fancy}
\fancyhf{}
\fancyhead[C]{\scriptsize Accepted by International Symposium on Networks, Computers and Communications (ISNCC): Next Generation Networking and Internet, ©2025 IEEE}
\renewcommand{\headrulewidth}{0pt}

\section{Introduction}

The rapid advancement of wireless communication technologies has expanded beyond terrestrial systems, extending into challenging environments such as aerial, space, and underwater domains, where high-speed and reliable data transmission is increasingly demanded \cite{BILEN2022101724}, \cite{9520342}. Sixth-generation (6G) communication systems aim not only to increase bandwidth but also to enable multi-layered goals such as AI-assisted routing, ultra-low latency, high reliability, and environmental adaptability. In this context, underwater communication emerges as a critical component of the 6G vision.

\begin{figure}[htbp]
    \centering
    \includegraphics[width=0.35\textwidth]{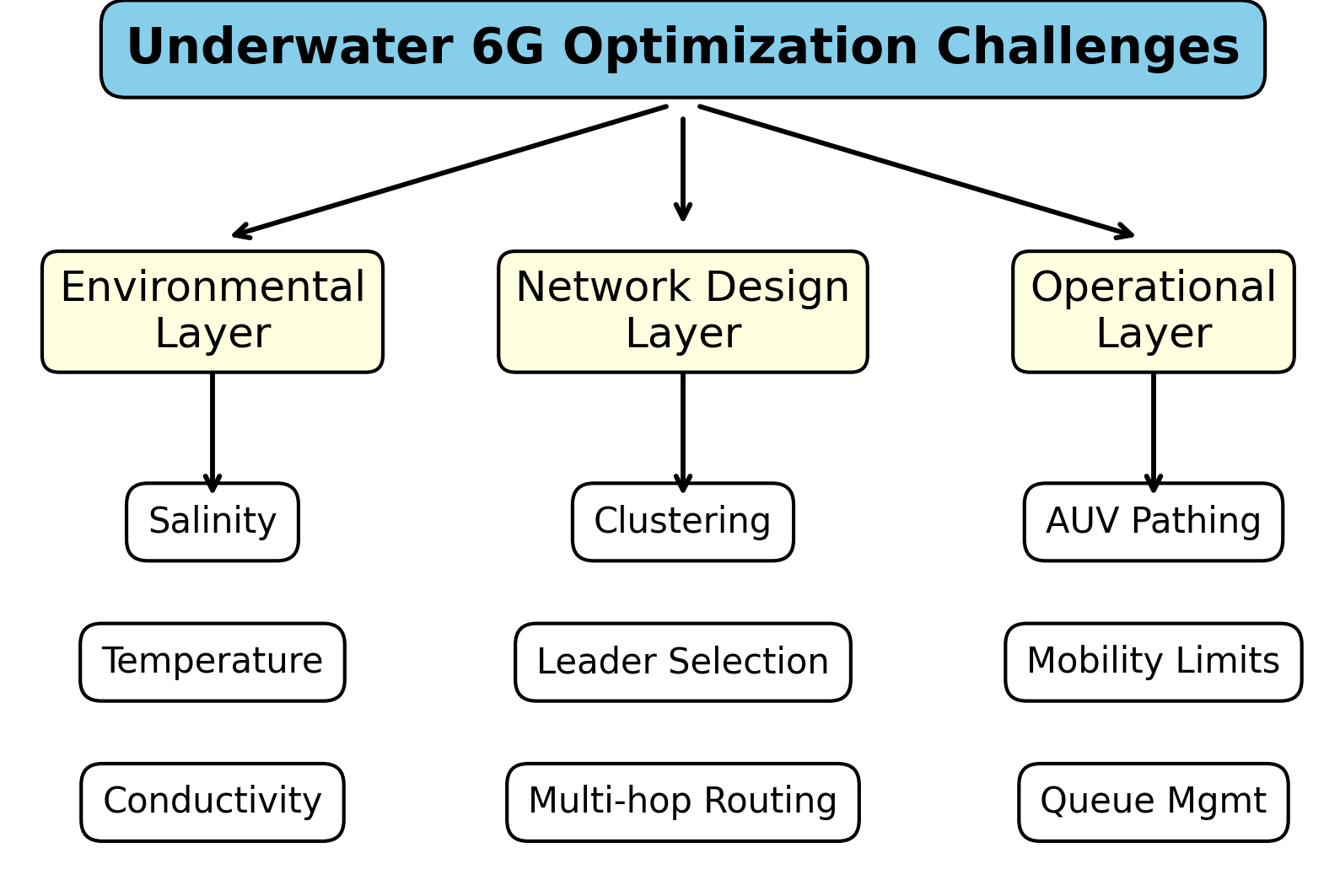}
    \caption{Challenges for Optimizing Underwater 6G Networks}
    \label{fig:challenges}
\end{figure}

However, the underwater environment includes complex physical and chemical parameters that significantly affect the propagation of electromagnetic (EM) signals. Variables such as salinity, temperature, pressure (depth), and conductivity lead to signal attenuation and reduced propagation speed, as summarized in Fig. \ref{fig:challenges}. Traditional acoustic communication systems are constrained by low frequency and limited data rates, whereas EM-based approaches offer high-speed transmission over short ranges. This study addresses these limitations by modeling a 6G communication network suitable for underwater electromagnetic propagation.

The proposed network architecture consists of fixed-position hub stations, mobile autonomous underwater vehicles (AUVs), and numerous static sensor nodes. The spatial configuration of each component is modeled in conjunction with the physical properties of seawater. Signal attenuation, transmission delay, and delivery success rate are calculated based on physics-based equations. The EM propagation model is further supported by parameters such as seawater conductivity and relative permittivity.

To improve overall network performance, a multi-stage optimization approach has been adopted. First, sensor nodes are grouped using the K-Means clustering algorithm \cite{9617810}. Then, each cluster undergoes sequential optimization via Genetic Algorithm (GA) and Particle Swarm Optimization (PSO) to determine optimal node placement. Cluster leaders are selected based on AUV and hub proximity and serve as primary carriers in multi-hop data forwarding. Unlike strict assignment models, each leader communicates with the most suitable AUV based on proximity and signal quality, allowing for dynamic load balancing across the network.

To evaluate the flexibility of the communication architecture, two scenarios have been developed: one based on cluster leaders (leader-based) and the other using direct communication without leaders (leaderless). Both configurations are comparatively analyzed based on performance metrics such as success rate, transmission losses, and delay. Additionally, to test the sustainability of the optimization under environmental variation, the system allows temperature and salinity levels to be dynamically adjusted and re-optimized accordingly.

The developed simulation framework includes queue management, traffic load monitoring, and link reliability analysis. It is implemented with a PyQt5-based real-time graphical interface, allowing users to visualize and compare different scenarios. As summarized in Fig. \ref{fig:contributions}, this study provides a comprehensive simulation approach for designing and optimizing underwater 6G networks in an environmentally adaptive manner.

\subsection{Related Work}

The study of underwater wireless sensor networks (UWSNs) has evolved significantly, particularly in the context of energy-efficient data transmission, electromagnetic wave propagation, and optimization-assisted routing.

\subsubsection{Electromagnetic Communication in Seawater}

Che et al.~\cite{che} and Al-Shamma'a et al.~\cite{shaw} conducted pioneering works evaluating the feasibility of electromagnetic (EM) wave propagation in seawater. Their studies emphasized the limitations of acoustic and optical communication in turbid and shallow environments, and promoted EM as a short-range, high-bandwidth alternative. These findings were experimentally validated by Shaw et al.~\cite{toal}, highlighting signal behavior under varying conductivity and salinity levels. Zahedi et al.~\cite{zahedi} further explored the feasibility of RF-EM communication, proposing its potential use for near-field underwater scenarios.

\subsubsection{Clustering and Routing Techniques}

Efficient clustering in underwater networks has been a focus of several studies. Li et al.~\cite{li} proposed the LE-KCR protocol, combining energy awareness and K-means clustering to reduce node death and improve longevity. Zhang et al.~\cite{sun} introduced an improved K-means-based clustered routing protocol tailored for underwater environments. Sun et al.~\cite{han} presented an adaptive clustering strategy that optimizes route selection under dynamic environmental conditions. Cho et al.~\cite{cho} offered a broader perspective, assessing the readiness of 6G underwater infrastructure and emphasizing the need for intelligent and resilient routing mechanisms.

\subsubsection{AUV-Assisted Data Collection and Localization}

Jing et al.~\cite{jing} introduced an AUV-assisted data collection strategy, optimizing path planning to minimize energy consumption. Saha and Arya~\cite{saha} integrated Particle Swarm Optimization (PSO) with virtual anchor nodes to enhance localization accuracy in dynamic underwater environments.

\subsubsection{Optimization-Based Deployment}

Optimization algorithms have been extensively applied to enhance network topology and communication reliability. Rao and Vijay~\cite{rao} leveraged Genetic Algorithms (GA) to determine optimal sensor placements. Mahmutoğlu et al.~\cite{yigit} utilized PSO to train a decision feedback equalizer for acoustic signal demodulation, improving error rates in noisy underwater channels.

\begin{figure}[htbp]
    \centering
    \includegraphics[width=0.38\textwidth]{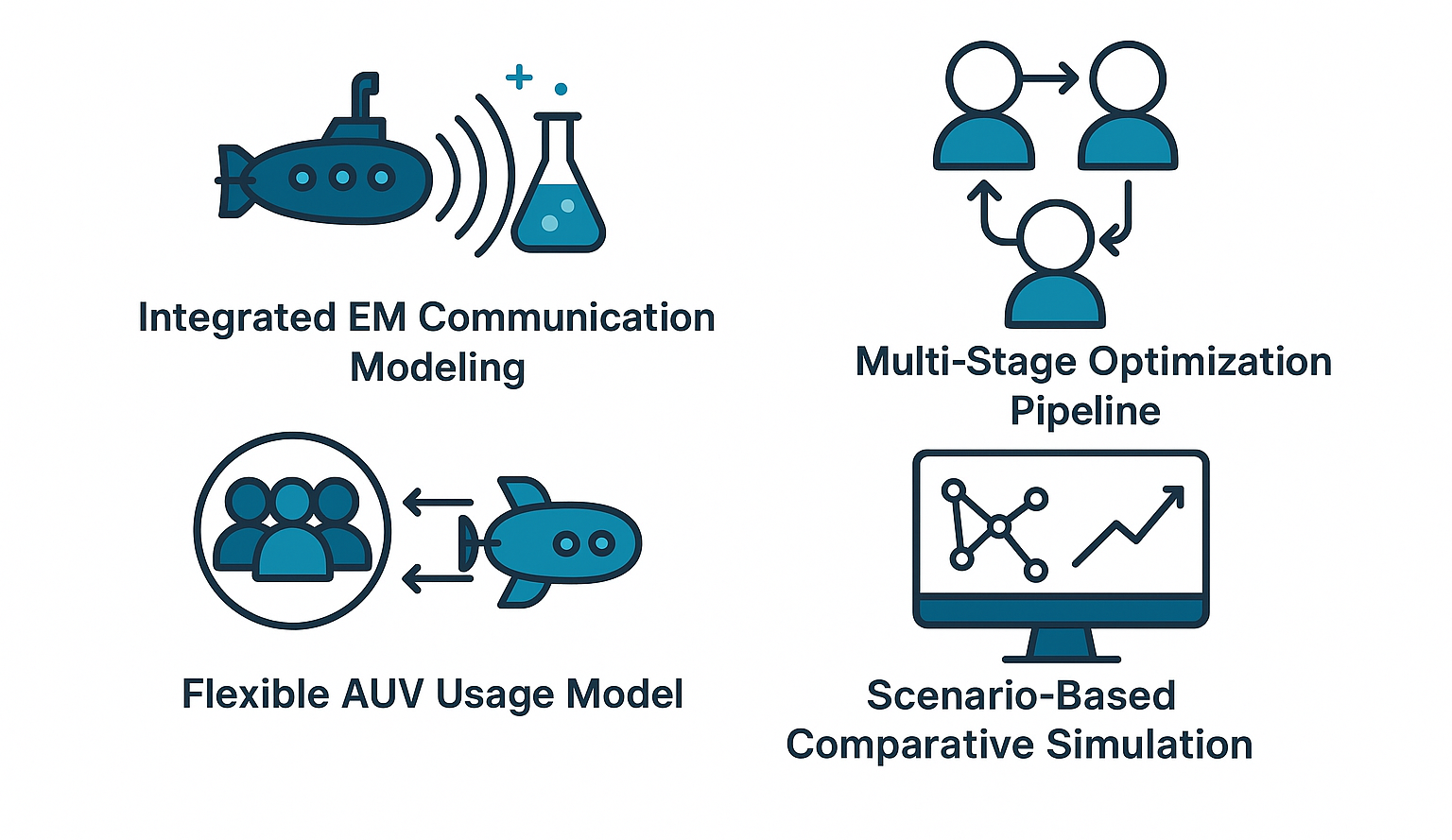}
    \caption{Main Contributions}
    \label{fig:contributions}
\end{figure}

Although these studies have explored underwater EM propagation, clustering-based routing, and AUV-assisted architectures, as illustrated in Fig. \ref{fig:contributions}, no prior work has unified these under a dynamic and adaptive simulation platform that integrates following details:

\begin{itemize}
    \item \textbf{Integrated EM Communication Modeling:} Unlike traditional acoustic-focused studies, this work leverages electromagnetic (EM) wave propagation with seawater-specific attenuation models (e.g., conductivity, temperature, salinity) to reflect more realistic underwater conditions.
    
    \item \textbf{Multi-Stage Optimization Pipeline:} A hybrid pipeline combining K-Means clustering, Genetic Algorithm (GA), and Particle Swarm Optimization (PSO) is sequentially applied to optimize the positions of sensor nodes, cluster leaders, and AUVs. This layered approach outperforms single-method techniques in terms of signal reliability and delivery ratio.
    
    \item \textbf{Flexible AUV Usage Model:} Unlike rigid role assignments in existing models, this framework allows cluster leaders to communicate with the nearest available AUV regardless of their own cluster affiliation. This dynamic allocation improves load balancing and reduces communication delays.
    
    \item \textbf{Scenario-Based Comparative Simulation:} The platform includes both leader-based and leaderless communication modes to evaluate trade-offs in terms of delay, packet success, and traffic load. This dual-scenario structure enables robust comparative analysis.
\end{itemize}

\section{The Proposed System Model}
\begin{figure}[htbp]
    \centering
    \includegraphics[width=0.33\textwidth]{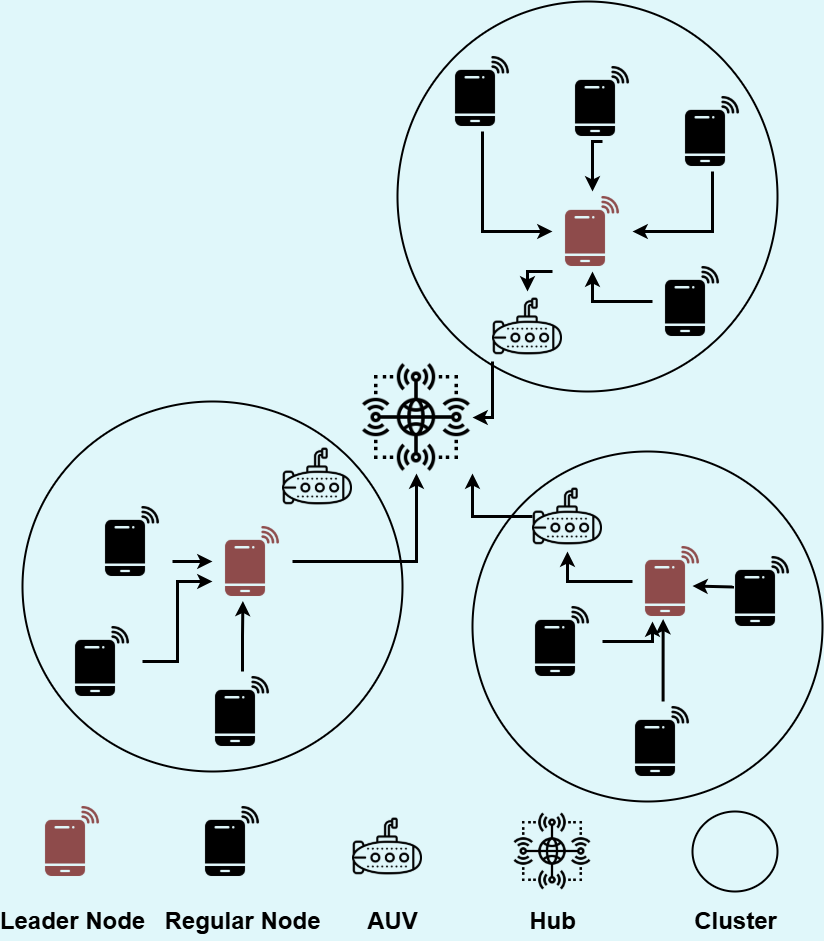}
    \caption{The structure of Leader-Based optimization model.}
    \label{fig:structure}
\end{figure}

This section presents the architecture and components of the proposed underwater 6G communication system. The system is designed to enable efficient electromagnetic (EM) communication in harsh underwater environments by dynamically adapting to environmental factors and optimizing node deployment \cite{10410731}. As illustrated in Fig.~\ref{fig:structure}, the network architecture is composed of the following components:

\begin{itemize}
    \item \textbf{Sensor Nodes:} Static nodes responsible for data generation and initial transmission \cite{7300803}. These nodes are grouped into clusters based on spatial proximity using K-Means clustering. Each cluster is further divided into two roles:
    \begin{itemize}
        \item \textit{Cluster Leaders:} One node per cluster is selected as the leader based on its proximity to the cluster centroid. Leaders collect data from cluster members and forward it to the hub either directly or via an AUV.
        \item \textit{Non-Leader Nodes:} In the Leader Node scenario, these nodes send their data to their respective cluster leader, rather than communicating directly with the hub. In the No-Leader scenario, these nodes send their data to the hub directly or via an AUV.
    \end{itemize}
    \item \textbf{Clusters:} Clusters are dynamically formed groups of sensor nodes that simplify network topology and reduce redundant transmissions. Each cluster operates semi-independently, with its own designated leader responsible for managing data forwarding on behalf of the group.
    \item \textbf{Autonomous Underwater Vehicles (AUVs):} Mobile relay nodes used for intermediate data collection and forwarding. AUVs are strategically positioned between cluster centers and hubs to improve link quality and mitigate long-range transmission losses.
    \item \textbf{Hubs:} Fixed-position stations that receive the final aggregated data and serve as gateways to upper-layers.
\end{itemize}

In this system model, each sensor node transmits a fixed number of packets either directly to the nearest hub (if within range) or via multi-hop routes using AUVs and cluster leaders. Signal attenuation and propagation delay are computed based on realistic EM wave models, accounting for frequency and medium properties. Also, in this model, environmental parameters such as temperature, salinity, and electrical conductivity significantly affect EM signal propagation. As detailed in the upcoming parts, these parameters are incorporated into the signal attenuation and speed formulas. This allows the communication range and success probability to vary dynamically with environmental changes.

\section{The Proposed Approach}

The proposed system adopts a hierarchical multi-stage optimization strategy to enhance the efficiency and reliability of underwater electromagnetic (EM) communication networks. This approach addresses the challenges of node deployment, signal attenuation, and delay in a dynamic marine environment by integrating unsupervised learning with evolutionary optimization techniques.

The pipeline begins with the application of the \textbf{K-Means clustering algorithm}, which partitions sensor nodes into $K$ spatially coherent clusters based on their Euclidean distances. Clustering not only reduces network complexity but also lays the foundation for localized coordination. Within each cluster, the sensor node that is closest to the cluster centroid is selected as the \textit{cluster leader}. These leaders act as local data aggregators, collecting packets from their respective cluster members and forwarding them either directly to a hub or via an intermediate Autonomous Underwater Vehicle (AUV).

Following the clustering stage, the system applies a \textbf{Genetic Algorithm (GA)} to optimize the physical locations of all nodes, including sensors and AUVs. GA leverages the principles of natural selection and uses operations such as tournament selection, crossover and mutation to iteratively evolve the node positions. The optimization objective is to minimize both intra-cluster distances and each node's distance to the nearest hub, ensuring strong connectivity and reduced energy consumption. This stage provides an efficient but approximate solution to the node placement problem.

To further refine this result, a \textbf{Particle Swarm Optimization (PSO)} algorithm is employed. PSO models the movement of a swarm of particles in the search space, where each particle adjusts its trajectory based on personal and global best experiences. Unlike GA, which emphasizes exploration, PSO focuses on fine-tuning by exploiting known good solutions. In this phase, the optimization function incorporates not only geometric distances but also environmental parameters such as signal attenuation and transmission delay. This ensures that the final node configuration is both energy-aware and latency-optimized under realistic underwater conditions.

By sequentially combining these three methodologies for clustering, genetic evolution, and swarm intelligence achieves a balance between local optimization (within clusters) and global coordination (across the entire network). The resulting network topology supports robust, scalable, and low-latency communication that adapts effectively to the underwater environment.

\subsection{Clustering with K-Means}

The optimization process begins with \textbf{K-Means clustering}, which groups sensor nodes into $K$ spatial clusters to simplify the network structure. Each cluster operates semi-independently, enabling localized optimization. The algorithm minimizes intra-cluster variance, as given in Eq. \ref{e1}

\begin{equation}\label{e1}
    J = \sum_{i=1}^{K} \sum_{x \in C_i} \| x - \mu_i \|^2
\end{equation}

In Eq. \ref{e1}, $C_i$ is the set of nodes in cluster $i$ and $\mu_i$ is its centroid. After clustering, one \textbf{Cluster Leader} is selected per cluster—specifically, the node closest to the centroid, as given in Eq. \ref{e2}.

\begin{equation}\label{e2}
    \text{Leader}_i = \arg\min_{x \in C_i} \| x - \mu_i \|
\end{equation}

These leaders collect data from their cluster and forward it to AUVs or hubs. The clustering output serves as the foundation for the GA and PSO stages.

\subsection{Node Placement Optimization with Genetic Algorithm}

In this study, a \textbf{Genetic Algorithm (GA)} based approach is adopted to optimize the spatial placement of each sensor node. GA is a heuristic search algorithm inspired by the process of natural evolution and employs operators such as \textit{selection}, \textit{crossover}, and \textit{mutation} to converge towards optimal solutions in the search space.

Here, the objective is to determine the optimal $(x_i, y_i)$ coordinates for $n$ sensor nodes in a 2D Euclidean space $\mathbb{R}^2$. The optimization seeks to minimize the total cost function as given in Eq. \ref{e3}.

\begin{equation}\label{e3}
  \min_{x_1, y_1, \dots, x_n, y_n} \left( \sum_{i=1}^{n} \sum_{\substack{j=1 \\ j \ne i}}^{n} d_{ij} + \sum_{i=1}^{n} \min_{k} d_{ih_k} \right)
\end{equation}

In Eq. \ref{e3}, $d_{ij} = \sqrt{(x_i - x_j)^2 + (y_i - y_j)^2}$ denotes the Euclidean distance between sensor nodes $i$ and $j$. The $d_{ih_k}$ is the distance between sensor $i$ and hub $k$. Also, $\min_{k} d_{ih_k}$ represents the minimum distance from each sensor to the nearest hub. This objective balances both \textbf{intra-network connectivity} and \textbf{external access to hubs}, promoting strong inter-node communication and low-latency access to data collection points.

In the proposed genetic algorithm, each individual is represented as a real-valued vector of length $2n$, encoding the $x$ and $y$ coordinates of $n$ nodes, i.e., $\text{individual} = [x_1, y_1, x_2, y_2, \dots, x_n, y_n]$. The initial population is generated randomly using a uniform distribution $\mathcal{U}(0, 100)$ to ensure coverage over the simulation area. For reproduction, the Blend Crossover (BLX-$\alpha$) operator is used to interpolate between pairs of parent individuals and create diverse offspring. Mutation is performed using Gaussian perturbation, which adds normally distributed noise to each coordinate with a predefined probability, introducing variability into the population.

To drive selection pressure, tournament selection is applied, favoring individuals with lower total cost while maintaining diversity. The algorithm operates over 40 generations with an initial population size of 30 individuals. Throughout this evolution process, the fitness of each individual is evaluated, and genetic operations are applied iteratively. At the end of the process, the individual yielding the minimum total cost is selected as the optimal solution.

After optimization, the position of each sensor node is obtained as $\text{Node } i \rightarrow (x_i, y_i) \quad \text{for } i = 1, 2, \dots, n$.

\subsection{Node Placement Optimization with Particle Swarm Optimization}

To further refine the placement of sensor nodes and improve overall communication efficiency, \textbf{Particle Swarm Optimization (PSO)} is employed following the Genetic Algorithm stage. PSO is a population-based stochastic optimization technique inspired by the social behavior of bird flocking or fish schooling. It explores the solution space by updating particles' velocities and positions based on personal and global bests.

Similar to the GA formulation, the goal is to determine optimal coordinates $(x_i, y_i)$ for $n$ sensor nodes that minimize the overall cost function, as given in Eq. \ref{e4}.

\begin{equation} \label{e4}
   \min_{x_1, y_1, \dots, x_n, y_n} \left( \sum_{i=1}^{n} \sum_{\substack{j=1 \\ j \ne i}}^{n} d_{ij} + \sum_{i=1}^{n} \min_{k} d_{ih_k} \right)
\end{equation}

In Eq. \ref{e4}, $d_{ij}$ and $d_{ih_k}$ are defined as in the GA section. The function ensures minimal intra-node distances and shortest access to the nearest hubs.

In the proposed Particle Swarm Optimization (PSO) algorithm, each particle encodes a candidate solution as a real-valued vector representing the $x$ and $y$ coordinates of $n$ nodes as $\text{particle} = [x_1, y_1, x_2, y_2, \dots, x_n, y_n]$. During each iteration, the velocity of a particle is updated according to Eq. \ref{e5}.
\begin{equation}\label{e5}
    v_i^{(t+1)} = w v_i^{(t)} + c_1 r_1 (p_i^{\text{best}} - x_i^{(t)}) + c_2 r_2 (g^{\text{best}} - x_i^{(t)})
\end{equation}

In Eq. \ref{e5}, $w$ denotes the inertia weight that balances exploration and exploitation, $c_1$ and $c_2$ are the acceleration coefficients that respectively control the influence of cognitive and social behavior, and $r_1$, $r_2$ are uniformly distributed random numbers in $[0, 1]$. The terms $p_i^{\text{best}}$ and $g^{\text{best}}$ refer to the personal best position of particle $i$ and the global best position found by the entire swarm, respectively. Once velocities are updated, the particle positions are adjusted using with Eq. \ref{e6}.
\begin{equation}\label{e6}
    x_i^{(t+1)} = x_i^{(t)} + v_i^{(t+1)}
\end{equation}


After optimization, the final node positions are obtained from PSO as $\text{Node } i \rightarrow (x_i, y_i) \quad \text{for } i = 1, 2, \dots, n$.

In addition to these, the objective function incorporates real-world environmental factors by modeling both signal attenuation and propagation delay as functions of seawater temperature and salinity. This makes the optimization highly relevant for practical deployment scenarios in dynamic underwater conditions.

\subsection{Signal Attenuation and Propagation Delay Modeling}

Underwater communication is significantly influenced by environmental parameters such as temperature, salinity, and conductivity. To model these effects realistically, both signal attenuation and propagation delay are incorporated into the model.

\subsubsection*{Attenuation Model}

The attenuation of electromagnetic (EM) waves in seawater is modeled based on an empirical approximation proposed in~\cite{qureshi}. Using vacuum permeability ($\mu_0$), the signal attenuation coefficient $\alpha$ (in dB/m) is calculated as given in Eq. \ref{e7}.

\begin{equation} \label{e7}
    \alpha(f, T, S) = 8.686 \cdot \sqrt{f \cdot \mu_0 \cdot \sigma(T, S) \cdot \pi}
\end{equation}

In Eq. \ref{e7}, $\sigma(T, S)$ denotes the electrical conductivity of seawater and is estimated as given in Eq. \ref{e8}.

\begin{equation} \label{e8}
    \sigma(T, S) = 0.19 \cdot S \cdot (1 + 0.02 \cdot (T - 25))
\end{equation}

Here, $T$ denotes the seawater temperature in degrees Celsius, and $S$ represents the salinity in practical salinity units (PSU).

Accordingly, the total attenuation over a transmission distance $d$ could be found with Eq. \ref{e9}.

\begin{equation}\label{e9}
    L(f, d, T, S) = \alpha(f, T, S) \cdot d
\end{equation}

Additionally, to reflect the stronger transmission capability of AUVs, the effective communication distance is scaled by a factor of 0.79 when the transmitter is an AUV, thereby reducing the total attenuation accordingly. Also, a fixed carrier frequency of $100\,\text{kHz}$ is used in all transmission scenarios to reflect realistic short-range underwater EM communication conditions~\cite{wang2019}.

\subsubsection*{Propagation Modeling and Cost Integration}

To realistically capture the physical behavior of electromagnetic communication underwater, both signal attenuation and propagation delay are considered in the cost function. The \textit{propagation delay} $\tau$ between two nodes separated by distance \(d\) is calculated as $\tau = \frac{d}{v_p}
\quad \text{where} \quad v_p \approx \frac{2f}{\mu \sigma}$. The \textit{phase velocity} \(v_p\) of electromagnetic waves in seawater is approximated by the relation described in \cite{wang2019}. For standard seawater conditions (\(\sigma \approx 4 \, \text{S/m}\), \(\mu = \mu_0\)) and a frequency of \(f = 100\,\text{kHz}\), the resulting velocity is approximately \(v_p \approx 5 \times 10^5 \, \text{m/s}\). This is significantly faster than acoustic propagation (typically \(1500 \, \text{m/s}\)), allowing lower latency in short-range underwater EM communication. Both \textit{attenuation} ($L$) and \textit{delay} ($\tau$) are jointly minimized in the PSO objective to achieve energy-efficient and delay-tolerant node placement. The overall cost is computed as $\text{Cost} = \alpha \cdot \sum L + \beta \cdot \sum \tau$. Here, $\alpha$ and $\beta$ are weighting coefficients that balance signal reliability and latency.

As a result, the proposed optimization strategy combines unsupervised learning and evolutionary algorithms in a hierarchical pipeline, formally expressed as $x^* = \text{PSO} \left( \text{GA} \left( \text{KMeans}(X) \right) \right)$. This structure ensures both local refinement and global optimization, enabling effective node placement under dynamic underwater constraints.

\section{Performance Evaluation}

\subsection{Simulation Environment}
To enable the execution of simulation experiments, a Python-based standalone \texttt{.exe} application was developed. The application was tested on a system running Windows 11, equipped with 16 GB of RAM and an Intel Core i7 processor. The developed software allows the user to input key parameters such as the number of sensor nodes, autonomous underwater vehicles (AUVs), and hubs, as well as environmental variables including water temperature and salinity. At the initialization stage of the simulation, the positions of all nodes and hubs are randomly generated. This stochastic placement strategy enables the evaluation of diverse spatial configurations, thus contributing to a more generalized assessment of system performance. Due to the high attenuation of electromagnetic (EM) waves at high frequencies in underwater environments, the simulation area is limited to a 100$\times$100 meter field. This constraint ensures realistic modeling while keeping nodes and AUVs within effective communication range. Specifically, the simulation framework is composed of three main scenarios:
\begin{itemize}
    \item \textbf{Initial Scenario:} Only direct communication between sensor nodes and hubs is allowed. AUVs remain inactive at this stage. This scenario serves as a baseline for performance comparison.
    \item \textbf{Leader-Based Scenario:} After optimization, each cluster designates a leader node. These leaders aggregate data from their respective cluster members and forward the aggregated packets to the hub, either directly or via AUVs.
    \item \textbf{Leaderless Scenario:} All sensor nodes are treated equally and transmit their data to the most suitable destination (either a hub or an AUV) based on signal quality and distance metrics.
\end{itemize}

In these scenarios, three modes of data routing are implemented:
\begin{itemize}
    \item \textbf{Initial:} Sensor → Hub directly (no AUVs, no clustering)
    \item \textbf{Leaderless:} Sensor → AUV/Hub (whichever is closer). All nodes act independently.
    \item \textbf{Leader-Based:} Sensor → Leader → AUV/Hub. Leaders are chosen per cluster and aggregate traffic.
\end{itemize}

Additionally, a custom simulator has been developed in Python using PyQt5. It supports real-time animation of packet flows, node communication, queue states, and network congestion. Moreover, to simulate realistic packet delivery behavior under signal attenuation, a logistic model is used. The delivery success probability is defined as $P_{\text{delivery}}(L) = \frac{1}{1 + e^{s \cdot (L - \theta)}}$. Here, $L$ is the total signal attenuation (in dB), $s$ is the slope parameter controlling sensitivity to attenuation, and $\theta$ is the threshold attenuation level at which the success rate is approximately 50\%. This probabilistic model captures the exponential decay in packet delivery as attenuation increases. It is integrated into the simulation to evaluate realistic success rates in both direct and multi-hop transmission scenarios.

\begin{figure*}[ht]
    \centering
    \subfloat[Average success rates across 100 simulation runs for each scenario.]{\label{fig:results}%
        \includegraphics[width=0.35\textwidth]{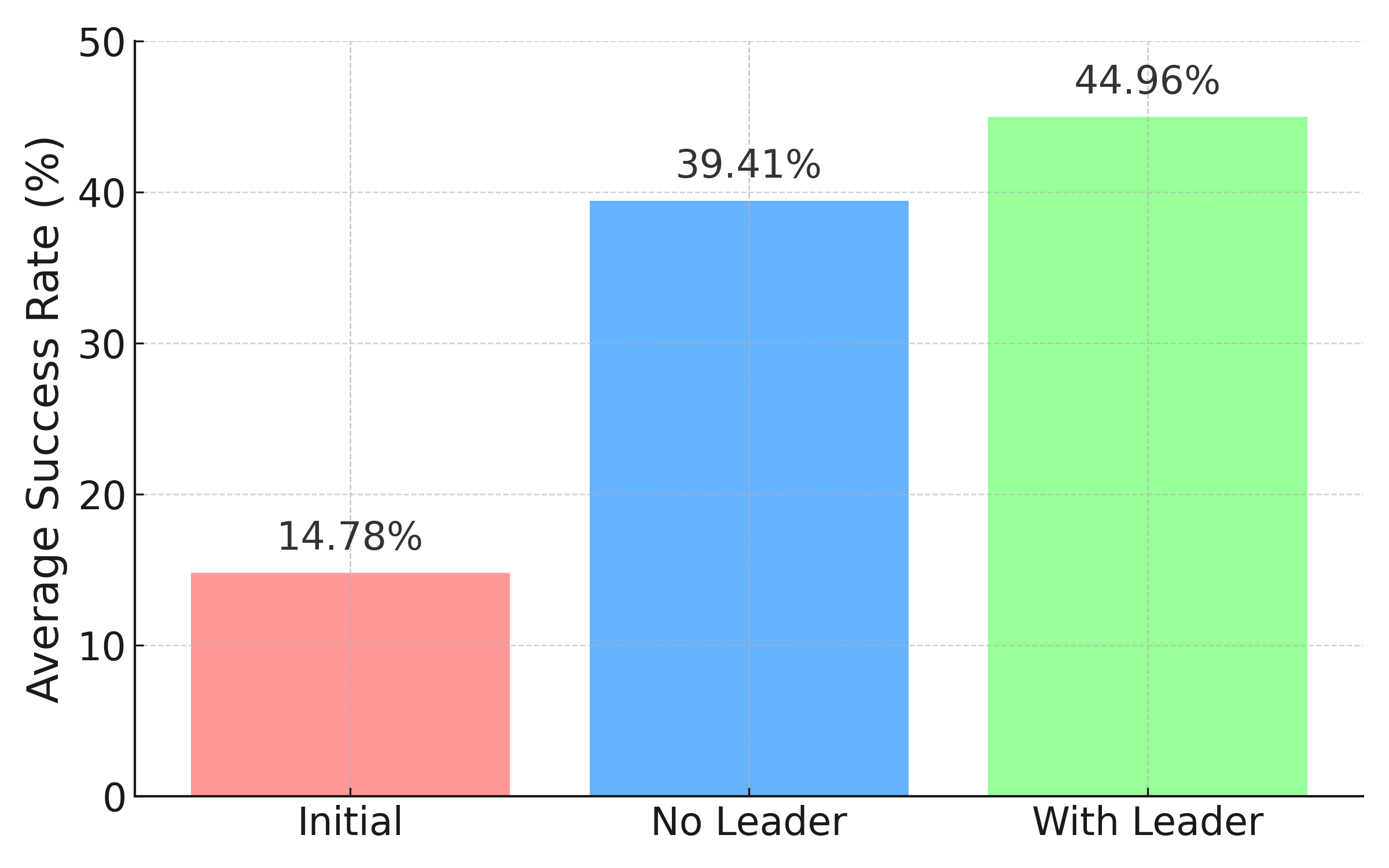}%
    }\hspace{0.6in}
    \subfloat[AUV Usage Rate by Communication Scenario.]{\label{fig:auv_usage}%
        \includegraphics[width=0.4\textwidth]{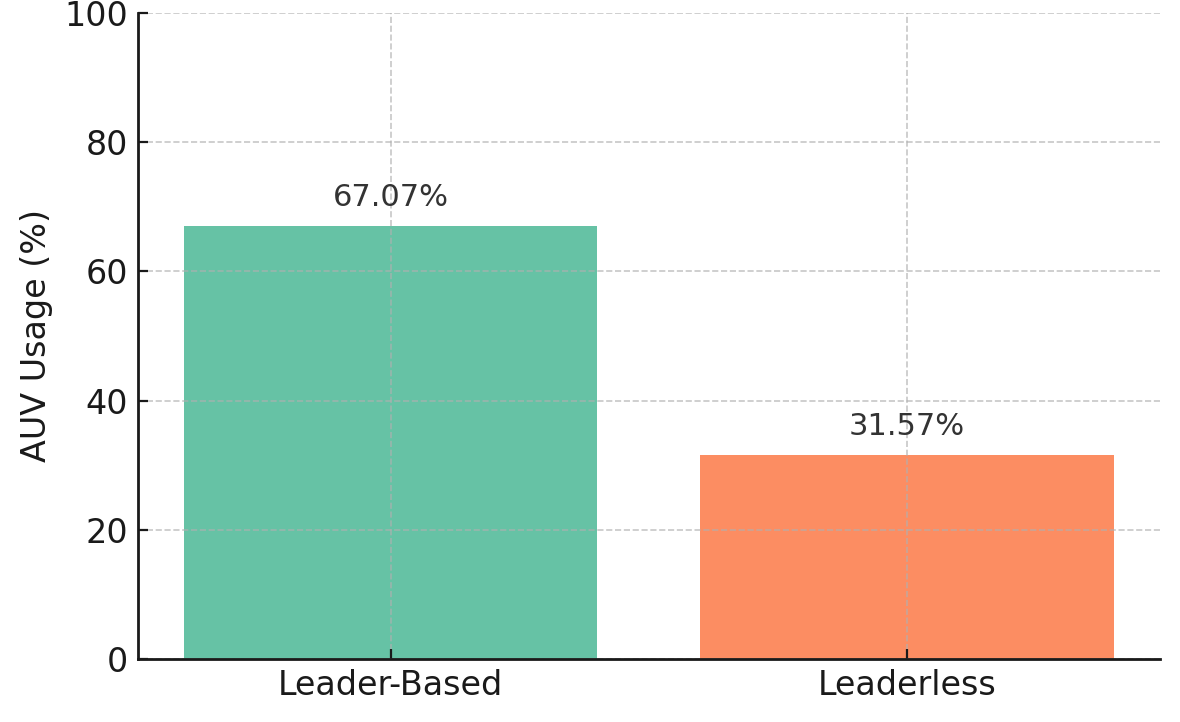}%
    }
    \caption{Evaluation results.}
    \label{r6}
\end{figure*}

\subsection{Simulation Results}
\begin{table}[htbp]
\centering
\tiny
\caption{Sample Routing Paths from a Simulation Run}
\label{tab:routes}
\begin{tabular}{|c|c|c|c|c|c|c|}
\hline
\textbf{Scenario} & \textbf{Node ID}   & \textbf{Route} & \textbf{Success} \\
\hline
Initial     & Sensor 0  & Direct → Hub 1        & 1.0 \\
Initial     & Sensor 1  & Direct → Hub 0        & 0 \\
Initial     & Sensor 2  & Direct → Hub 0        & 0 \\
Initial     & Sensor 3  & Direct → Hub 0        & 0 \\
Initial     & Sensor 4  & Direct → Hub 0        & 0 \\
Initial     & Sensor 5  & Direct → Hub 0        & 0 \\
Initial     & Sensor 6  & Direct → Hub 1        & 1.0 \\
Initial     & Sensor 7  & Direct → Hub 1        & 1.0 \\
Initial     & Sensor 8  & Direct → Hub 0        & 0 \\
Initial     & Sensor 9  & Direct → Hub 1        & 1.0 \\
\hline
No-Leader   & Sensor 0  & AUV 2 → Hub 1         & 1.0 \\
No-Leader   & Sensor 1  & AUV 1 → Hub 1         & 1.0 \\
No-Leader   & Sensor 2  & Direct → Hub 1        & 1.0 \\
No-Leader   & Sensor 3  & Direct → Hub 0        & 0.49 \\
No-Leader   & Sensor 4  & AUV 4 → Hub 0         & 0 \\
No-Leader   & Sensor 5  & Direct → Hub 0        & 0 \\
No-Leader   & Sensor 6  & Direct → Hub 0        & 0 \\
No-Leader   & Sensor 7  & AUV 2 → Hub 1         & 1.5 \\
No-Leader   & Sensor 8  & Direct → Hub 0        & 0 \\
No-Leader   & Sensor 9  & AUV 1 → Hub 1         & 1.0 \\
\hline
Leader      & 0         & AUV 2 → Hub 1         & 1.0 \\
Leader      & 1         & AUV 1 → Hub 1         & 1.0 \\
Leader      & 4         & AUV 4 → Hub 0         & 0 \\
Leader      & 3         & Direct → Hub 0        & 0.49 \\
Leader      & 7         & AUV 2 → Hub 1         & 1.5 \\
\hline
\end{tabular}
\end{table}
Table~\ref{tab:routes} presents a snapshot from a single simulation instance, illustrating the routing paths chosen in each scenario. Each sensor node transmits 50 packets. The table highlights which route was used (via AUV or directly to hub) and whether the transmission was successful. As observed in Table~\ref{tab:routes}, AUVs are actively utilized in both the leader and no-leader scenarios, whereas the \textit{initial scenario} exclusively employs direct transmissions from sensor nodes to hubs, with all AUVs being inactive. This variability highlights the importance of route-aware optimization and adaptive relay selection to enhance overall communication reliability.

Based on these, to obtain statistically robust and interpretable results, 100 independent simulations were conducted, each with distinct initial random placements. The performance of each scenario was individually recorded and averaged across all runs. Given the inherent variability of each random configuration, this averaging approach captures the general trends and mitigates the impact of outliers. As shown in Figure~\ref{fig:results}, these findings demonstrate that the optimization procedures significantly enhance communication reliability. Although the leader-based scenario generally achieves the highest average success rate, the leaderless scenario occasionally outperforms it in certain configurations due to its adaptive nature. In rare cases, the initial scenario yields comparable or better performance, especially when randomly placed nodes are already favorably aligned with hubs.

To understand the impact of communication topology on AUV load, we calculated the percentage of transmissions that utilize AUVs in both leader-based and leaderless scenarios. As shown in Fig. \ref{fig:auv_usage}, the results of average 100 random simulations show a significantly higher usage rate in the leader-based scenario, where 67.07\% of all successful transmissions involved AUVs, compared to 31.57\% in the leaderless setup. This disparity stems from the routing structure as illustrated in Table \ref{tab:routes}. Specifically, in leader-based communication, leaders typically act as aggregators and forward large volumes of data through AUVs. In contrast, in the leaderless configuration, nodes are more likely to send data directly to hubs when possible, reducing reliance on AUV relays.

\begin{figure}[htbp]
    \centering
    \includegraphics[width=0.4\textwidth]{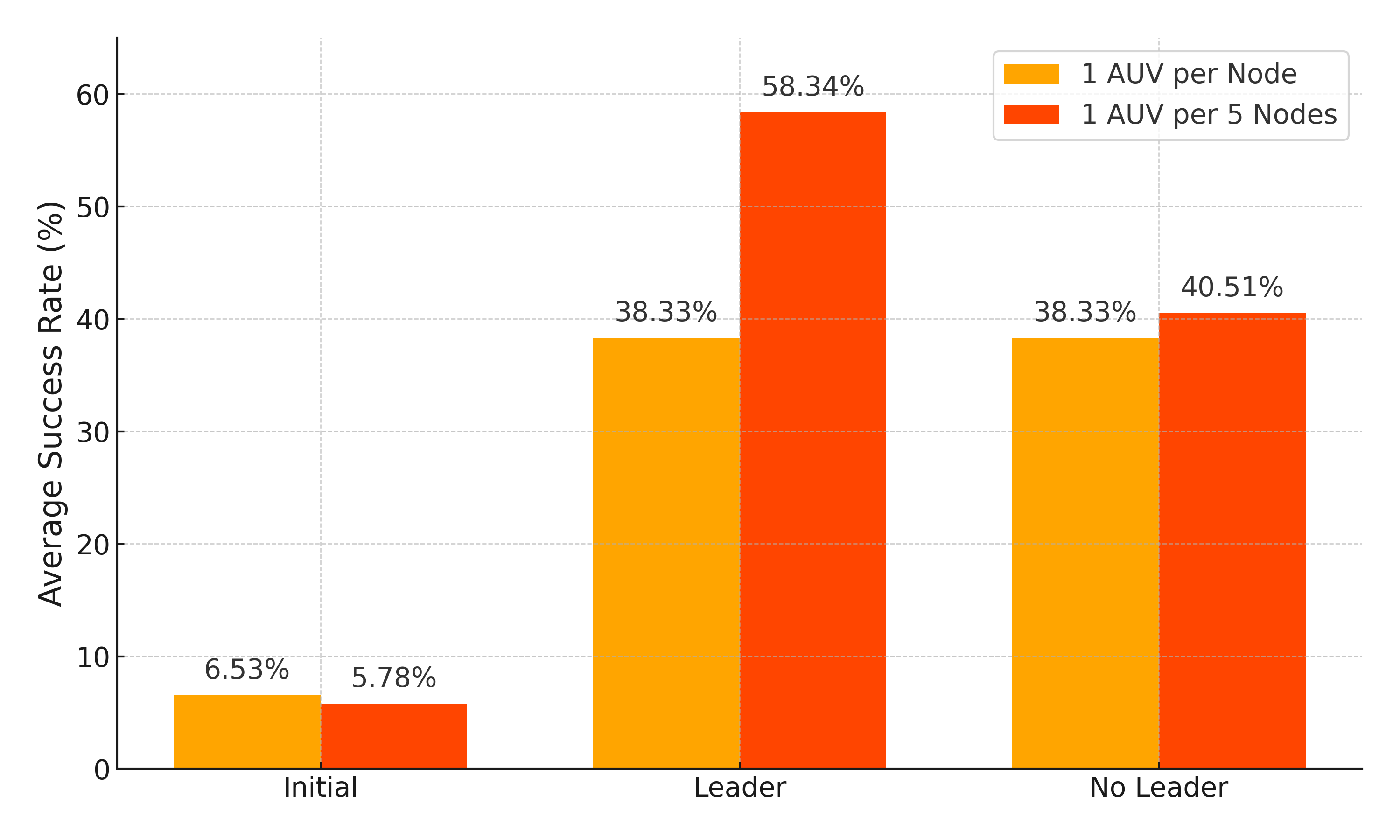}
    \caption{Scenario-Wise Success Rate By AUV Assignment}
    \label{fig:auv_success}
\end{figure}

Additionally, to evaluate the impact of AUV density on communication performance, two different configurations were compared: one AUV assigned per sensor node (1 AUV per node) and one AUV for every five sensor nodes (1 AUV per 5 nodes). Each scenario (Initial, Leader-Based, and No-Leader) was executed 100 independent simulations with same number of hubs, same temperature, same salinity and average success rates were calculated as illustrated in Fig.~\ref{fig:auv_success}. In the Initial scenario, since no optimization is applied and AUVs remain passive, only direct transmission to hubs is performed, resulting in a notably low success rate. In the configuration where there is one AUV per sensor, each cluster consists of a single node, causing every sensor to become a leader by default. Consequently, the Leader-Based and No-Leader scenarios yield the same result. When AUV density is reduced, a significant increase in success rate is observed in the Leader-Based scenario. This seemingly counterintuitive outcome suggests that limiting the number of AUVs can actually improve data transmission efficiency by reducing network congestion and collisions when data is routed via cluster leaders. In the No-Leader scenario, the change in AUV quantity does not result in a major shift in success rate, further demonstrating the effectiveness of the applied optimization for node positioning. Overall, the results show that AUV density does not always have a linear effect on communication success, and in some cases, excessive AUV deployment may even lead to inefficiencies in network performance.

It is important to emphasize that the results of each simulation are inherently subject to variability due to the random initialization of node and hub positions, as well as probabilistic transmission success. Therefore, identical outcomes across all simulation runs cannot be guaranteed. Instead, the reported averages reflect general performance trends rather than deterministic behavior. Overall, the leader-based architecture offers the most consistent performance gains by improving data aggregation and transmission efficiency in clustered underwater networks.

\section{Conclusion}
This study presents a comprehensive simulation framework for evaluating electromagnetic (EM) communication in underwater 6G sensor networks. By integrating clustering, optimization, and realistic signal attenuation models, the system allows for the assessment of different data transmission strategies under varying environmental and topological conditions. The simulation supports three main communication scenarios: a baseline initial setup, a leaderless strategy, and a leader-based hierarchical model. Results obtained from 100 randomized trials indicate that optimized scenarios significantly outperform the baseline, with the leader-based approach demonstrating the highest average success rate. These findings highlight the importance of data aggregation and intelligent routing in mitigating underwater communication challenges such as signal attenuation and transmission unreliability. While the current system captures key aspects of short-range EM underwater communication, it also lays the groundwork for more advanced features such as energy modeling, dynamic mobility, and learning-based optimization. In conclusion, the developed simulation tool not only validates the feasibility of EM-based underwater communication strategies but also provides a scalable and extensible base for future research in next-generation underwater networks.

\section{Future Work}

Future enhancements will focus on increasing realism and scalability. Energy consumption models will be added to evaluate power-performance trade-offs, and mobility support for AUVs and sensor nodes will reflect real-world dynamics. Reinforcement learning methods such as DQN or PPO may replace heuristic optimizers for adaptive decision-making. GUI improvements like signal heatmaps and interactive controls are also planned. Finally, large-scale simulations with hundreds of nodes will assess the system’s robustness under high network loads.

\bibliographystyle{IEEEtran}
\bibliography{references}

\end{document}